# Analytical Bounds on Maximum-Likelihood Decoded Linear Codes with Applications to Turbo-Like Codes: An Overview

Igal Sason, Shlomo Shamai

Technion - Israel Institute of Technology, Haifa 32000, Israel. E-mails: {sason, sshlomo}@ee.technion.ac.il

## Abstract

Upper and lower bounds on the error probability of linear codes under maximum-likelihood (ML) decoding are shortly surveyed and applied to ensembles of codes on graphs. For upper bounds, focus is put on Gallager bounding techniques and their relation to a variety of other reported bounds. Within the class of lower bounds, we address de Caen's based bounds and their improvements, sphere-packing bounds, and information-theoretic bounds on the bit error probability of codes defined on graphs. A comprehensive overview is provided in the monograph [22].

## 1 Introduction

Consider the classical coded communication model of transmitting one of equally likely signals over a communication channel. Since the error performance of coded communication systems rarely admits exact expressions, tight analytical upper and lower bounds emerge as a useful theoretical and engineering tool for assessing performance and for gaining insight into the effect of the main system parameters. As specific good codes are hard to identify, the performance of ensembles of codes is usually considered. The Fano [9] and Gallager [12] bounds were introduced as efficient tools to determine the error exponents of the ensemble of random codes, providing informative results up to the ultimate capacity limit. Since the advent of information theory, the search for efficient coding systems has motivated the introduction of efficient bounding techniques tailored to specific codes or some carefully chosen ensembles of codes. A classical example is the adaptation of the Fano upper bounding technique [9] to specific codes, as reported in the seminal dissertation by Gallager [11] (to be referred to as the 1961 Gallager-Fano bound). The incentive for introducing and applying such bounds has strengthened with the introduction of various families of codes defined on graphs which closely approach the channel capacity with feasible complexity (e.g., turbo codes, repeat-accumulate codes, and low-density parity-check (LDPC) codes). Clearly, the desired bounds must not be subject to the union bound limitation, since for long blocks these ensembles of turbo-like codes perform reliably at rates which are considerably above the cutoff rate ($R_0$) of the channel (recalling that union bounds for long codes are not informative at the portion of the rate region above $R_0$, where the performance of these capacity-approaching codes is most appealing). Although maximum-likelihood (ML) decoding is in general prohibitively complex for long codes, the derivation of upper and lower bounds on the ML decoding error probability is of interest, providing an ultimate indication of the system performance. Further, the structure of efficient codes is usually not available, necessitating efficient bounds on performance to rely only on basic features, such as the distance spectrum and the input-output weight enumeration function (IOWEF) of the examined code (for the evaluation of the block and bit error probabilities, respectively, of a specific code or ensemble). These latter features can be found by analytical methods (see e.g., [18]).

In [22], we present various reported upper bounds on the ML decoding error probability. These include the bounds of Berlekamp [2], Divsalar [5], Duman-Salehi [6], Engdahl-Zigangirov [8], Gallager-Fano 1961 bound [11], Hughes [15], Poltyrev [19], Sason-Shamai [21], Shulman-Feder [28], Viterbi ([30] and [31]), Yousefi-Khandani ([33] and [34]) and others. We demonstrate in [22] the underlying connections that exist between them; the bounds are based on the distance spectrum or the IOWEFs of the codes. The focus of this presentation is directed towards the application of efficient bounding techniques on ML decoding performance, which are not subjected to the deficiencies of the union bounds and therefore provide useful results at rates reasonably higher than the cut-off rate. In [22] and references therein, improved upper bounds are applied to block codes and turbo-like codes. In addressing the Gallager bounds and their variations, we focus in [25] (and more extensively in [22, Chapter 2]) on the Duman and Salehi variation which originates from the standard Gallager bound. A large class of efficient recent bounds (or their Chernoff versions) is demonstrated to be a special case of the generalized second version of the Duman and Salehi bounds. Implications and applications of these observations are addressed in Section 3.

We also address here lower bounds on the ML decoding error probability (see [22, Chapter 3]), and exemplify these bounds on linear block codes with a special emphasis on ensembles of codes defined on graphs. Specifically, we overview a class of bounds which are based on de Caen's bound and an improved version of it, sphere-packing bounds and some recent developments on these bounds, and other information-theoretic lower bounds which are suitable for linear codes defined on bipartite graphs (e.g., LDPC codes).

## 2 General Approach for the Derivation of Improved Upper Bounds

In [22, Chapter 2], we present many improved upper bounds on the ML decoding error probability which are tighter than the union bound. Let $\mathbf{r}$ be the received signal vector, and $\mathcal{R}$ be an arbitrary region around the transmitted signal point. The basic concept which is common to the derivation of the upper bounds within the class discussed in [22, Chapter 2] is the following:

$$\begin{aligned} \Pr(\text{error}) &= \Pr(\text{error}, \mathbf{r} \in \mathcal{R}) + \Pr(\text{error}, \mathbf{r} \notin \mathcal{R}) \\ &\leq \Pr(\text{error}, \mathbf{r} \in \mathcal{R}) + \Pr(\mathbf{r} \notin \mathcal{R}). \quad (1) \end{aligned}$$

The region $\mathcal{R}$ in (1) is interpreted as the "good region". The idea is to use the union bound only for the joint event where the decoder fails to decode correctly, and in addition, the received signal vector falls inside the region $\mathcal{R}$ (i.e., the union bound is used for upper bounding the first term in the right-hand side (RHS) of (1)). On the other hand, the second term in the RHS of (1) represents the probability of the event where the received signal vector falls outside the region $\mathcal{R}$, and which is typically the dominant term for very low SNR, is calculated only one time (and it is not part of the event where the union bound is used). We note that in the case where the region $\mathcal{R}$ is the whole observation space, the basic approach which is suggested above provides the union bound. However, since the upper bound in (1) is valid for an arbitrary region $\mathcal{R}$ in the observation space, many improved upper bounds can be derived by an appropriate selection of this region. These bounds could be therefore interpreted as geometric bounds (see [5] and [25]). As we will see, the choice of the region $\mathcal{R}$ is very significant in this bounding technique; different choices of this region have resulted in various different improved upper bounds which are considered extensively in [22, Chapter 2]. For instance, the tangential bound of Berlekamp [2] used the basic inequality in (1) to provide a considerably tighter bound than the union bound at low SNR values. This was achieved by separating the radial and tangential components of the Gaussian noise with a half-space as the underlying region $\mathcal{R}$. For the derivation of the sphere bound [13], Herzberg and Poltyrev have chosen the region $\mathcal{R}$ in (1) to be a sphere around the transmitted signal vector, and optimized the radius of the sphere in order to get the tightest upper bound within this form. Divsalar bound [5] is another simple and tight bound which relies on the basic inequality (1). The geometrical region $\mathcal{R}$ in Divsalar bound was chosen to be a sphere; in addition to the optimization of the radius of this sphere, the center of the sphere which does not necessarily coincide with the transmitted signal vector was optimized too. Finally, the tangential-sphere bound (TSB) which was proposed for binary linear block codes by Poltyrev [19] and for M-ary PSK block coded-modulation schemes by Herzberg and Poltyrev [14] selected $\mathcal{R}$ as a circular cone of half-angle $\theta$, whose central line passes through the origin and the transmitted signal. It is one of the tightest upper bounds known to-date for linear codes whose transmission takes place over a binary-input AWGN channel (see Fig. 1 and [20], [21], [33]).

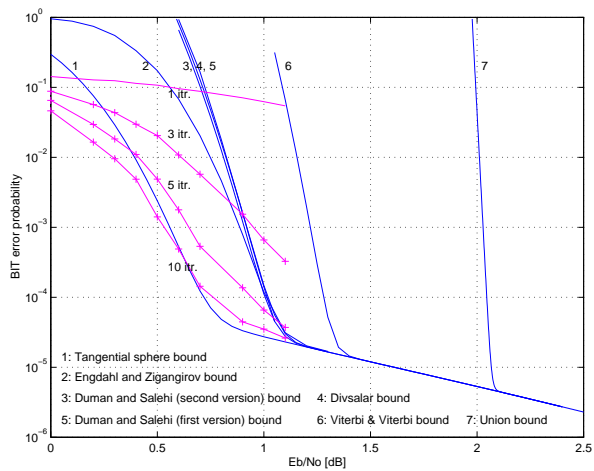

Fig. 1. Various bounds for the ensemble of rate-$\frac{1}{3}$ turbo codes whose components are recursive systematic convolutional codes with generators $G_1(D) = G_2(D) = \left[1, \frac{1+D^4}{1+D+D^2+D^3+D^4}\right]$. There is no puncturing of the parity bits, and the uniform interleaver between the two parallel concatenated (component) codes is of length $N = 1000$. It is assumed that the transmission of the codes takes place over a binary-input AWGN channel. The upper bounds on the bit error probability under optimal ML decoding are compared with computer simulations of the iterative sum-product algorithm with up to 10 iterations.

We note that the bounds mentioned above are only a sample of various bounds reported in [22, Chapter 2]; all of these bounds rely on the inequality (1) where the geometric region $\mathcal{R}$ characterizes the resulting upper bounds on the decoding error probability. After providing the general approach, we outline some connections between these bounds and demonstrate a few possible applications.

## 3 On Gallager Bounds: Variations and Applications

In addressing the Gallager bounding techniques and their variations, we focus in [22] and [25] on a generalization of the second version of the Duman and Salehi (DS2) bound [7]. The Duman and Salehi bounding technique (see [7] and [25]) originates from the

1965 Gallager bound [12]. Let $\psi_N^m(\underline{y})$ be an arbitrary probability measure (which may also depend on the transmitted codeword $\underline{x}^m$). The 1965 Gallager bound then yields that for arbitrary $\lambda, \rho > 0$:

$$P_{e|m} \leq \sum_{\underline{y}} \psi_N^m(\underline{y}) \, \psi_N^m(\underline{y})^{-1} \, p_N(\underline{y}|\underline{x}^m) \left( \sum_{m' \neq m} \left( \frac{p_N(\underline{y}|\underline{x}^{m'})}{p_N(\underline{y}|\underline{x}^m)} \right)^\lambda \right)^\rho$$

$$= \sum_{\underline{y}} \psi_N^m(\underline{y}) \left( \psi_N^m(\underline{y})^{-\frac{1}{\rho}} \, p_N(\underline{y}|\underline{x}^m)^{\frac{1}{\rho}} \sum_{m' \neq m} \left( \frac{p_N(\underline{y}|\underline{x}^{m'})}{p_N(\underline{y}|\underline{x}^m)} \right)^\lambda \right)^\rho.$$

By invoking Jensen's inequality to the RHS of the last inequality, the DS2 bound follows:

$$P_{e|m} \leq \left( \sum_{m' \neq m} \sum_{\underline{y}} p_N(\underline{y}|\underline{x}^m)^{\frac{1}{\rho}} \, \psi_N^m(\underline{y})^{1-\frac{1}{\rho}} \left( \frac{p_N(\underline{y}|\underline{x}^{m'})}{p_N(\underline{y}|\underline{x}^m)} \right)^\lambda \right)^\rho$$

for arbitrary parameters $0 \leq \rho \leq 1$ and $\lambda \geq 0$.

Let $G_N^m(\underline{y})$ be an arbitrary non-negative function of $\underline{y}$, and let the probability density function $\psi_N^m(\underline{y})$ be

$$\psi_N^m(\underline{y}) = \frac{G_N^m(\underline{y}) \, p_N(\underline{y}|\underline{x}^m)}{\sum_{\underline{y}} G_N^m(\underline{y}) \, p_N(\underline{y}|\underline{x}^m)}. \quad (2)$$

The functions $G_N^m(\underline{y})$ and $\psi_N^m(\underline{y})$ are referred to as the *un-normalized and normalized tilting measures*, respectively. The substitution of (2) into the RHS of the last inequality gives

$$P_{e|m} \leq \left( \sum_{\underline{y}} G_N^m(\underline{y}) \, p_N(\underline{y}|\underline{x}^m) \right)^{1-\rho}$$
$$\cdot \left\{ \sum_{m' \neq m} \sum_{\underline{y}} p_N(\underline{y}|\underline{x}^m) \, G_N^m(\underline{y})^{1-\frac{1}{\rho}} \left( \frac{p_N(\underline{y}|\underline{x}^{m'})}{p_N(\underline{y}|\underline{x}^m)} \right)^\lambda \right\}^\rho$$

where $0 \leq \rho \leq 1$ and $\lambda \geq 0$. This upper bound on the decoding error probability was also derived in [5, Eq. (62)]. For the class of memoryless channels, and for the choice of $\psi_N^m(\underline{y})$ as $\psi_N^m(\underline{y}) = \prod_{i=1}^N \psi^m(y_i)$ (recalling that the function $\psi^m$ may also depend on the transmitted codeword $\underline{x}^m$), this upper bound is relatively easily evaluated (similarly to the standard union bounds) for particular block codes. In that case, the latter bound is calculable in terms of the distance spectrum, not requiring the fine details of the code structure. A similar upper bound on the bit error probability is expressible in terms of the IOWEFs of the codes (or the average IOWEFs of code ensembles).

By generalizing the framework of the DS2 bound, a large class of efficient bounds (or their Chernoff versions) is demonstrated to follow from this bound. Implications and applications of these observations are pointed out in [25], including the fully interleaved fading channel, resorting to either matched or mismatched decoding. The proposed approach can be generalized to geometrically uniform non-binary codes, finite state channels, bit-interleaved coded-modulation systems, parallel channels [17], and it can be also used for the derivation of upper bounds on the conditional decoding error probability. In [22] and [25], we present the suitability of variations on the Gallager bounds as bounding techniques for random and deterministic codes, which partially rely on insightful observations made by Divsalar [5]. Focus is put in [25] on geometric interpretations of the 1961 Gallager-Fano bound (see [9] and [11]). The interconnections between many reported upper bounds are illustrated in [25, Fig. 8], where it is shown that the generalized DS2 bound particularizes to these upper bounds by proper selections of the tilting measure. Further details, extensions and examples are provided in [22].

The TSB [19] happens often to be the tightest reported upper bound for block codes which are transmitted over the binary-input AWGN channel and ML decoded (see e.g., [20] and [21]). However, in the random coding setting, it fails to reproduce the random coding error exponent (see [19]), while the DS2 bound does. In fact, also the Shulman-Feder bound [28] which is a special case of the latter bound achieves capacity for the ensemble of fully random block codes. This substantiates the claim that there is no uniformly best bound. However, we note that the loosened version of the TSB [5] (which involves the Chernoff inequality) maintains the asymptotic (i.e, for infinite block length) exponential tightness of the TSB of Poltyrev [19], and it is a special case of the DS2 bound.

In the following, we exemplify the use of the DS2 bounding technique for fully interleaved fading channels with faulty measurements of the fading samples.

*Example 1: The Generalized DS2 bound for the Mismatched Regime.* In [25], we apply the generalized DS2 bound to study the robustness of a mismatched decoding that is based on ML decoding with respect to the faulty channel measurements. We examine here the robustness of the decoder in case that a BPSK modulated signal is transmitted through a fully interleaved Rayleigh fading channel. For simplicity, the bounds are applied to the case of a perfect phase estimation of the i.i.d fading samples (in essence reducing the problem to a real channel). We also assume here that the estimated and real magnitudes of the Rayleigh fading samples have a joint distribution of two correlated bivariate Rayleigh variables with an average power of unity.

The bounds in Fig. 2 refer to the ensemble of uniformly interleaved rate $-\frac{1}{3}$ turbo codes whose components are recursive systematic convolutional codes: $G_1(D) = G_2(D) = \left[1, \frac{1+D^4}{1+D+D^2+D^3+D^4}\right]$ without puncturing of parity bits, and an interleaver length of $N = 1000$. Since for a fully interleaved Rayleigh fading channel with *perfect* side information on the fading samples, the channel matched cutoff rate corresponds to $\frac{E_b}{N_0} = 3.23$ dB then, according to the upper bounds depicted in Fig. 2, the ensemble performance of these turbo codes (associated with the ML decoding) is sufficiently robust in case of mismatched decoding, even in a portion of the rate region exceeding the chan-nel matched cutoff rate. The proposed upper bounds depicted here were efficiently implemented in software, indicating their feasible computational complexity.

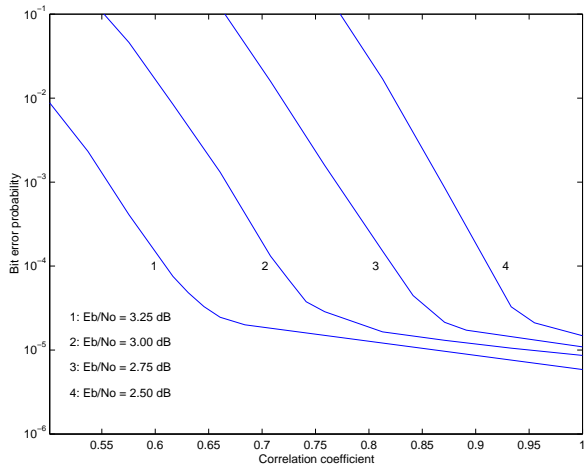

Fig. 2. A comparison between upper bounds on the bit error probability for the ensemble of turbo codes considered in Example 1 where the transmission of these codes takes place over a fully interleaved Rayleigh fading channel with mismatched decoding. The bounds are based on the combination of the generalized DS2 bound and the tight form of the union bound applied to every constant Hamming-weight subcode. These bounds are plotted for $\frac{E_b}{N_0} = 2.50, 2.75, 3.00$ and 3.25 dB, as a function of the correlation coefficient between the actual i.i.d Rayleigh fading samples and their Rayleigh distributed estimations.

# 4 Lower Bounds on the Decoding Error Probability

## 4.1 De Caen Inequality and Variations

D. de Caen [3] suggested a lower bound on the probability of a finite union of events. While an elementary result (essentially, Cauchy-Schwartz inequality), this bound was used to compute lower bounds on the decoding error probability of linear block codes via their distance distribution (see [16] for the binary symmetric channel (BSC), and [24] for the Gaussian channel). In [4], Cohen and Merhav improved de Caen's inequality by introducing a weighting function which is subject to optimization. Their improved bound is presented in the following statement and, like de Caen's inequality, it follows from the Cauchy-Schwartz inequality.

*Theorem 4.1:* [4, Theorem 2.1] Let $\{A_i\}_{i \in \mathcal{I}}$ be an arbitrary set of events in a probability space $(\Omega, \mathcal{F}, P)$, then the probability of the union of these events is lower bounded by

$$P\left(\bigcup_{i \in \mathcal{I}} A_i\right) \geq \sum_{i \in \mathcal{I}} \left\{ \frac{\left(\sum_{x \in A_i} p(x) m_i(x)\right)^2}{\sum_{j \in \mathcal{I}} \sum_{x \in A_i \cap A_j} p(x) m_i(x)^2} \right\} \quad (3)$$

where $m_i(x) \geq 0$ is any real function on $\Omega$ such that the sums on the RHS of (3) converge. Further, equality in (3) is achieved when

$$m_i(x) = m^*(x) = \frac{1}{\deg(x)}, \quad \forall\, i \in \mathcal{I} \quad (4)$$

where for each $x \in \Omega$

$$\deg(x) \triangleq |\{i \in \mathcal{I} \mid x \in A_i\}|. \quad (5)$$

The lower bound on the union of events in Theorem 4.1 particularizes to de Caen's inequality by the particular choice of the weighting functions $m_i(x) = 1$ for all $i \in \mathcal{I}$, which then gives:

$$P\left(\bigcup_{i \in \mathcal{I}} A_i\right) \geq \sum_{i \in \mathcal{I}} \frac{P(A_i)^2}{\sum_{j \in \mathcal{I}} P(A_i \cap A_j)^2}.$$

Cohen and Merhav relied on (3) for the derivation of improved lower bounds on the decoding error probability of linear codes under optimal ML decoding. They exemplified their bounds for BPSK modulated signals which are equally likely to be transmitted among $M$ signals, and the examined communication channels were a BSC and an AWGN channel. In this context, the element $x$ in the RHS of (3) is replaced by the received vector $\mathbf{r}$ at the output of the communication channel (whose length $K$ is equal to the dimension of the code), and $A_i$ (where $i = 1, 2, \ldots, M-1$) consists of all the vectors of length $K$ which are closer in the Euclidean sense to the signal $\mathbf{s}_i$ rather than the transmitted signal $\mathbf{s}_0$. Following [24], the bounds in [4] get (after some loosening in their tightness) final forms which solely depend on the distance spectrum of the code. Recently, two lower bounds on the ML decoding error probability of linear binary block codes were derived by Behnamfar et al. [1] for BPSK-modulated AWGN channels. These bounds are easier for numerical calculation, but are looser than Cohen-Merhav bounds for low to moderate SNRs.

Note that de Caen's based lower bounds on the decoding error probability (see [1], [4], [16] and [24]) are applicable for *specific* codes but not for ensembles; this restriction is due to the fact that Jensen's inequality does not allow to replace the distance spectrum of a linear code in these bounds by the average distance spectrum of ensembles.

## 4.2 Sphere-Packing Bounds Revisited for Moderate Block Lengths

In the asymptotic case where the block length of a code tends to infinity, the best known lower bound on the decoding error probability for high levels of noise is the 1967 sphere-packing (SP67) bound [27]. Like the random coding bound of Gallager [12], the sphere-packing bound decreases exponentially with the block length. Further, the error exponent of the SP67 bound is a convex function of the rate which is known to be tight at the portion of the rate region between the critical rate ($R_c$) and the channel capacity; for this important rate region, the error exponent of the SP67 bound coincides with the error exponent of the random coding bound [27, Part 1]. For the AWGN channel, the 1959 sphere-packing (SP59) bound was derived by Shannon [26] by first showing that the error probability of any code whose codewords lie on a sphere must be greater than the error probability of a code of the same block length and rate whose codewords are uniformly distributed over that sphere.

The reason that the SP67 bound fails to provide useful results for codes of small to moderate block length is due to the original focus in [27] on asymptotic analysis. In their paper [29], Valembois and Fossorier have recently revisited the SP67 bound in order to make it applicable for codes of moderate block lengths, and also to extend its field of application to continuous output channels (e.g, the AWGN channel which is the communication channel model of the SP59 bound of Shannon [26]). The motivation for the study in [29] was strengthened due to the outstanding performance of codes defined on graphs even of moderate block length. The remarkable improvement in the tightness of the SP67 was exemplified in [29] for the case of the AWGN channel with BPSK signaling, and it was shown that the improved version of the SP67 is an interesting alternative to the SP59 bound [26].

### 4.3 Lower Bounds for Codes on Graphs

In [23, Theorem 2.5], Sason and Urbanke derived an information-theoretic lower bound on the bit error probability for binary linear block codes under ML decoding. The bound is expressed in terms of the *density* of an arbitrary parity-check matrix which represents a binary linear block code, and it is valid for any memoryless binary-input output-symmetric channel. The motivation for the derivation of this bound was due to the fact that the decoding complexity per iteration and the parity-check density are strongly linked when such a binary linear code is represented by a bipartite graph and an iterative message-passing decoding algorithm is applied to decode this code. The general idea used for the derivation of the lower bound on the bit error probability in [23, Theorem 2.5] was the derivation of a lower bound on the normalized conditional entropy $\frac{H(\mathbf{X}|\mathbf{Y})}{n}$, on one hand, where $\mathbf{X}$ and $\mathbf{Y}$ are the transmitted codeword and the received vector, respectively, and $n$ designates the block length of the code; this lower bound on the conditional entropy was combined with the Fano inequality, in order to finally obtain a lower bound on the bit error probability. The bound in [23, Theorem 2.5] is expressed in terms of the *normalized density* of an arbitrary parity-check matrix which represents the binary linear code; this quantity is defined to be equal to $t = \left(\frac{R}{2-R}\right)\Delta$ where $R$ is the code rate (in bits per channel use) and $\Delta$ designates the number of ones per information bit in the parity-check matrix. The reason for the scaling factor of $\frac{R}{2-R}$ is due to the fact that by doing so, the case where $t = 1$ corresponds to a bipartite graph without cycles, and otherwise $t > 1$ (the fundamental cycles in the bipartite graph grows linearly with $t$ [23]). The tightness of the lower bounds was demonstrated in [23] for the binary erasure channel (BEC). Except for the BEC and BSC, the tightness of these bounds was recently improved by Wiechman and Sason [32]. The improved bound in [32] is based on improving the tightness of the lower bound on the conditional entropy; to this end, the derivation of the latter bound relies on the soft values of the log-likelihood ratio (LLR) at the output of the channel, instead of a two-level quantization of the LLR which was used in [23] (thus degrading the original memoryless binary-input output-symmetric channel into a BSC). Another refinement in the tightness of the resulting lower bound on the bit error probability was obtained by using the Fano inequality to get the inequality $\frac{H(\mathbf{X}|\mathbf{Y})}{n} \leq R h_2(P_b)$ where $R$ is the code rate and $h_2(x) = -x\log_2(x) - (1-x)\log_2(1-x)$ designates the binary entropy function to the base 2 (the factor $R$ in this upper bound was previously replaced by 1 in the derivation of [23, Theorem 2.5]).

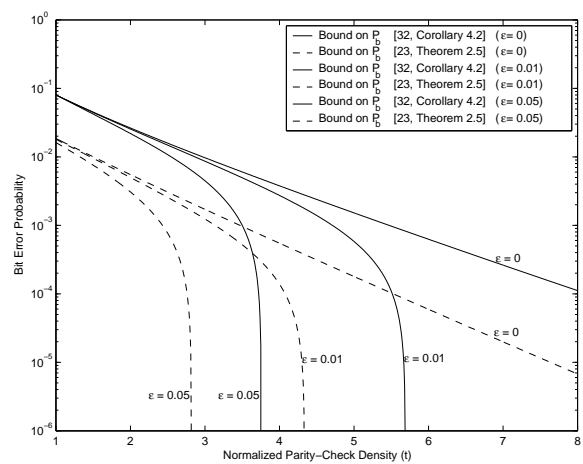

Fig. 3. Lower bounds on the bit error probability for any binary linear block code transmitted over a binary-input AWGN channel whose capacity is $\frac{1}{2}$ bits per channel use. The bounds are depicted in terms of the normalized density of an arbitrary parity-check matrix which represents the code, and the curves correspond to code rates which are a fraction $1-\varepsilon$ of the channel capacity (for different values of $\varepsilon$). The bounds depicted in dashed lines and solid lines are based on [23, Theorem 2.5] and [32, Corollary 4.2], respectively.

A lower bound on the BER is shown in Fig. 3 for the binary-input AWGN channel. These bounds are plotted as a function of the normalized density of an arbitrary parity-check matrix. In this example, the capacity of the channel is equal to $\frac{1}{2}$ bit per channel use, and the bounds are depicted for binary linear block codes whose rate is a fraction $1 - \varepsilon$ of the channel capacity. The plots in Fig. 3 show the superiority of the lower bound in [32, Eq. (78)] over the bound in [23, Theorem 2.5]. In order to exemplify the superiority of the former bound over the latter bound, assume for example that one wishes to design a binary LDPC code which achieves a bit-error probability of $10^{-6}$ at a rate which is $99\%$ of the channel capacity. The curve of the lower bound from [23] for $\varepsilon = 0.01$ implies that the normalized density of an arbitrary parity-check matrix which represents the code should be at least $4.33$, while the curve depicting the bound from [32, Eq. (78)] strengthens this requirement to a normalized density (of each parity-check matrix) of at least $5.68$. Translating this into terms of parity-check density (which is also the complexity per iteration for iterative message-passing decoding) yields minimal parity-check densities of $13.16$ and $17.27$, respectively (the minimal parity-check density is given by $\Delta_{\min} = \frac{(2-R)t_{\min}}{R}$). It is reflected from Fig. 3 that as the gap to capacity $\varepsilon$ tends to zero, the lower bound on the normalized density ($t$) of an arbitrary parity-check matrix grows significantly; it actually behaves like $\log \frac{1}{\varepsilon}$, and becomes unbounded as the gap to capacity vanishes.

**Acknowledgment**: The work was supported by the EU 6$^{\text{th}}$ International Framework Programme via the NEWCOM Network of Excellence.